# Yes and…? Using Improv to Design for Narrative in *Lights Out*

**Alina Striner**

Human-Computer interaction Lab

iSchool | University of Maryland

College Park, MD 20740, USA

algol001@umd.edu

## Abstract
Mixed-reality experiences often require detailed narrative that can be used to craft physical and virtual design components. This work elaborates on a mentoring experience at the Carnegie Mellon's ETC to consider how improv games may be used ideate and iterate on storytelling experiences.

## ACM Classification Keywords
HCI design and evaluation methods; Mixed / augmented reality; Interaction techniques

## Introduction
When animators render a scene in low lighting, they painstakingly model the details of the set in high fidelity, because even when viewers are unable to see them, the details allow the light to reflect more naturally than they would without the detail [11] (see Figure 1). As in animation, designing clear and specific narratives is vital to creating effective narrative experiences.  While traditional theater may use elaborate sets, costumes, and staging to communicate story [4], interactive performances often take place in diverse spaces [12], focusing on interaction in place of set [10]. Not obliged to think through the details of a narrative, mixed-reality designers may not generate the detail resolution that has the power to transform a satisfactory experience into a compelling one. Improv may be a tool that can help designers actively generate detailed story experiences and integrate story details into physical design spaces. This work describes and elaborates on a mentoring experience at the CMU Entertainment Technology Center (ETC) to consider how improv may be used as a narrative design method.

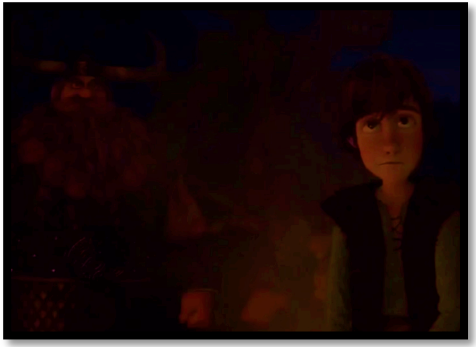

Figure 1: Scene from How To Train Your Dragon, one of the first films to use the low lighting technique described by Seymour [11].

## Related Work
In a well-designed environment, the blurred line between playing and making has the impetus to bring together designers from diverse backgrounds to share insights and skills during co-design. Games facilitate natural team building and promote the development of leadership skills through goal setting, interpersonal relations, and problem solving [2], and successful teams create trust by asking participants to role-play, perform collaboratively to complete a mission [3] or fulfill a narrative [5].

*Improv in Design*
Although collaborative games have been used to facilitate teamwork [5], improv games have not only been shown to increase cooperation and build trust, but have also been used in ideation and participatory design [7]. For instance, Simsarian [8] describes "bodystorming," the concept of brainstorming through role-play, and Svanæs [9] elaborates on using improve techniques in low-fidelity prototyping. Further, Medler intimately explores the cognition of improv actors to inform the design of synthetic characters [5]. HCI designers frequently use improv, however the medium has not yet been explored as a way of composing narrative and using it to shape interactions.

## The *Lights Out* Project
*Lights Out* was a nonvisual, multisensory, location-based interactive experience created by masters' students at the ETC as a semester design project [1]. During the experience, audiences felt their way around pitch-black space, solving auditory, olfactory, and touch puzzles.

*Integrating Narrative*
The project used a series of sensory experiments and play testing to design a fun and challenging integrated experience [1]. When I joined the project as a mentor, the students wanted to integrate the multisensory puzzles into a story, but were unsure of what narrative would fit with the story, and how to design the narrative into the puzzle experience.

*Step 1: Using Improv to Choose a Story*
Recognizing the need for an effective story, I asked students to generate narrative ideas; where and when the story took place, who the characters were, what their goals were, and what the puzzles represented in each narrative. After brainstorming 3 potential narratives, the students evaluated the effectiveness of each by playing the improv game, *Stage Directions* [3]; 2 team members took turns solving the puzzle tasks as the characters in each narrative, moving through the space and talking the way the characters might, while others supplied the actors with stage directions. For instance, after saying "lets explore the inside of this ship," a spaceman could be directed to use his instruments to check the ship's air quality, or remove his helmet. This game endowed each narrative with physicality, and created instinctive motivations for each puzzle interaction. After play through each scenario, the group discussed the merits of each narrative: what interesting dialogues and movements came out during the game, and how naturally each scenario fit the puzzles. At the end of the session, group members compared experiences; they chose a narrative that could be communicated most effectively in limited time.

*Outcomes* Students decided on a fantasy narrative where audience members played "mind-doctors." Using the multisensory puzzles, they were tasked to find and unlock a happy memory trapped inside Ebenezer Scrooge's unconscious mind.

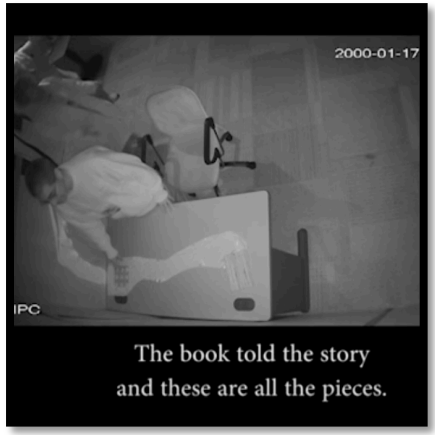

Figure 2: The audio puzzle represented Scrooge's diary [1].

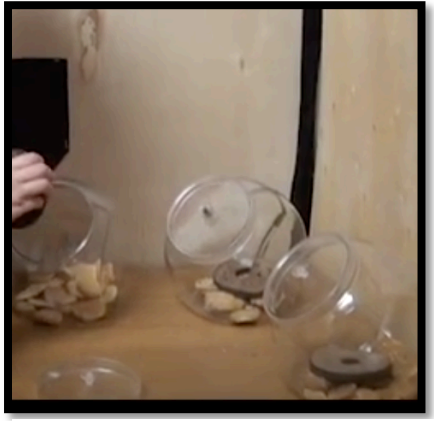

Figure 3: Students redesigned the spell puzzle for a set of kitchen cookie jars [1].

*Step 2: Defining Physical Space and Puzzle Meanings*
Having chosen a narrative, I asked students to more closely consider the design of the physical space and the meaning behind each puzzle interaction. *How did the narrative world map onto the physical space of the room?* This included considering what the unconscious mind of Scrooge looked like: *should the physical space resemble the inside of his head, or should the space represent a specific location?* This task built on the *Free Association* and *Freeze* improv games [3]; students were tasked to think of and act out interaction possibilities as they thought of them.

*Outcomes* Students decided that the physical space should represent Scrooge's childhood, and should be mapped to memories of his childhood home; a kitchen table where he smelled his mother's baking, a treehouse nook where young Scrooge hid his keepsakes, and a bedroom desk where he wrote in his diary. I also cued students to consider the temporal relationships of these memories; the kitchen memory would have been from early childhood, the treehouse was of early adolescence (when a child might hide things), and the desk represented Scrooge in his teens, grasping with maturity. In this way, audiences could navigate through the space temporally.

*Step 3: Using Narrative to Refine Interactions*
Armed with an understanding of the story and the representations of the physical space and puzzles tasks, I asked students to consider what specific interactions and movements represented, and how the physical environment could more fully reflect the Scrooge narrative. In this step, students focused on the beginning of the experience, considering how to transform audiences into the mind-doctor characters. To develop an effective introduction, students took turns role-playing as the narrator and audience participants, trying voice and gesture techniques from the Emotional Transfer improv game [3] to impart the urgency of the mind doctors onto audiences.

*Outcomes* After the improv design tasks, students had no trouble further integrating narrative into the story; they added physicality to the characters by giving participants lab coats to put as the experience began, and created a narrated transition space that clearly established the mind-doctor characters, the memory setting, and puzzle tasks. The students also defined the temporal space through musical cues (e.g. using a nursery melody to convey early childhood), and embedded narrative elements into each puzzle; for instance, they integrated the audio puzzle with Scrooge's diary memory, and fashioned buttons for the puzzle into an old book (Figure 2). Likewise, they redesigned the smell puzzle for a set of kitchen cookie jars (Figure 3).

## Discussion and Conclusions
In the *Light's Out* project, improv games were used to help students choose an narrative and refine the story experiences. Improvisation allowed students to actively move through the space – temporarily becoming characters and participants as a way to understand the many needs of the narrative. By creating a clear understanding of narrative, project designers were able to intuitively communicate the richness of the narrative through audience interactions; each of the puzzle tasks referenced a girl named "Rosie;" after solving all of the puzzles, participants unlocked Scrooge's happy memory, which was of her.

*Storytelling in a Non-Visual Medium*
Lights Out took place entirely in the dark, so the experience was freed from visual narratives conventions such as communicating time-period and space with set design [4]. This constraint freed the designers from having to fill the space with set items, and removed the complication of audience members disengaging from the narrative because of set quality. Since audience members were not able to judge the set visually, a few carefully curated set pieces conveyed the narrative more effectively than a furnished space.

## Future Work

Given the effectiveness of improv as a narrative generation tool, future work should consider what improv game tools would be most helpful to designers developing narrative-based experiences at different stages of design. *What improv games can help designers when they are working by themselves vs. in a group? What tools can help designers develop a story from scratch, elaborate upon an existing narrative, or integrate a story into an existing experience?*

Although the focus of this work was on interactive narrative, improv may also inform narrative design in a range of performances space such as choreography, film scoring, and costume design.

## Acknowledgements

Thank you to Heather Kelley for the mentorship opportunity, and to the team members of *Lights Out*, Alex Amorati, Dasol Park, Dale Wones, and Annie Wang.

## References

1. Anon. Light's Out. Retrieved February 7, 2017 from https://www.etc.cmu.edu/blog/projects/lights-out/
2. Dimitry Davidoff (1999). The Original Mafia Rules [Internet Archive]. Davidoff. Retrieved from http://web.archive.org/web/19990302082118/http://members.theglobe.com/mafia_rules/
3. Hugh Mcleod (2016). Teambuilding through Improvisation, 1–31. Retrieved from http://medianet-ny.com/TeamBuilding.pdf
4. Orville Kurth Larson. 1989. *Scene Design in the American Theatre from 1915 to 1960*. University of Arkansas Press.
5. Ben Medler and Brian Magerko. 2010. The implications of improvisational acting and role-playing on design methodologies. In *Proceedings of the SIGCHI Conference on Human Factors in Computing Systems* (CHI '10). ACM, New York, NY, USA, 483-492.
6. Nasir, M., Lyons, K., Leung, R., Bailie, A., and Whitmarsh. F. (2015). The effect of a collaborative game on group work. In *Proceedings of the 25th Annual International Conference on Computer Science and Software Engineering* (CASCON '15), Jordan Gould, Marin Litoiu, and Hanan Lutfiyya (Eds.). IBM Corp., Riverton, NJ, USA, 130-139.
7. Hilary O'Shaughnessy and Nicholas Ward. 2014. The use of physical theatre improvisation in game design. In *Proceedings of the 8th Nordic Conference on Human-Computer Interaction: Fun, Fast, Foundational* (NordiCHI '14).
8. Kristian T. Simsarian. 2003. Take it to the next stage: the roles of role playing in the design process. In *CHI '03 Extended Abstracts on Human Factors in Computing Systems* (CHI EA '03). ACM, New York, NY, USA, 1012-1013.
9. Dag Svanaes and Gry Seland. 2004. Putting the users center stage: role playing and low-fi prototyping enable end users to design mobile systems. In *Proceedings of the SIGCHI Conference on Human Factors in Computing Systems* (CHI '04). ACM, New York, NY, USA, 479-486.
10. Stuart Reeves, Christian Greiffenhagen, Martin Flintham, Steve Benford, Matt Adams, Ju Row Farr, and Nicholas Tandavantij. 2015. I'd Hide You: Performing Live Broadcasting in Public. In CHI '15.
11. Mike Seymour (2012). SIGGRAPH 2012 OPENS. Retrieved from: https://www.fxguide.com/featured/siggraph-2012-opens/
12. Keren Zaiontz. 2012. Ambulatory Audiences and Animate Sites: Staging the Spectator in Site-Specific Performance. In *Performing Site-Specific Theatre: Politics, Place, Practice*, Anna Birch and Joanne Tomkins. Palgrave Macmillan, New York, NY, 10010, 167-18